\documentclass[11pt]{article}
\usepackage{amsmath}
\usepackage{graphicx}

\newcommand{\xvec}{\mbox{\bf x}}
\newcommand{\Deltavec}{\mbox{\boldmath $\Delta$}}
\newcommand{\Xvec}{\mbox{\bf X}}

\newcommand{\sigmat}{\mbox{\boldmath $\Sigma$}}
\newcommand{\muvec}{\mbox{\boldmath $\mu$}}
\newcommand{\thetavec}{\mbox{\boldmath $\theta$}}

\begin{document}

\begin{center}
{\Large {\bf Test for the statistical significance of a treatment effect\\ in the presence of hidden sub-populations}}

\normalsize
{ B. Karmakar$^{+}$, K. Dhara$^{+}$, K.K. Dey$^{+}$, A. Basu$^{\circ}$, and A.K. Ghosh$^{+}$}

{\small
\vspace{0.0in}
$^{+}$ Indian Statistical Institute, 203, B. T. Road, Kolkata 700108, India.\\
$^{\circ}$ National Institute of Biomedical Genomics, Kalyani 741251, India.

\vspace{0.0in}
{\small Email: bikram.karmakar@yahoo.in, kumaresh.dhara@gmail.com,\\ kshldey@gmail.com, ab1@nibmg.ac.in, akghosh@isical.ac.in.}
}

\normalsize
\begin{abstract}
For testing the statistical significance of a treatment effect, we usually compare between two parts of a population, one is exposed to the treatment, and the other is not exposed to it. Standard parametric and nonparametric two-sample tests are often used for this comparison. But
direct applications of these tests can yield misleading results, especially when the population has some hidden sub-populations, and the impact of this sub-population difference on the study variables dominates the treatment effect. This problem becomes more evident if these sub-populations have widely different proportions of representatives in the samples taken from these two parts, which are often referred to as the treatment group and the control group. In this article, we make an attempt to overcome this problem. Our propose methods use suitable clustering algorithms to find the hidden sub-populations and then eliminate the sub-population effect by using suitable transformations. Standard two-sample tests, when they are applied on the transformed data, yield better results. Some simulated and real data sets are analyzed to show the utility of the proposed methods.

\vspace{0.1in}
\end{abstract}

\end{center}

\section{Introduction}

In a two-sample problem, we test the quality of the two distributions $F$ and $G$ based on two sets of independent observations $\xvec_{11},\xvec_{12},\ldots,\xvec_{1n_1}\stackrel{i.i.d.}{\large {\sim}} F$ and $\xvec_{21},\xvec_{22},\ldots,\xvec_{2n_2}\stackrel{i.i.d.}{\large {\sim}} G$. If $F$ and $G$ are assumed to be same except for their location, (i.e., $F(\xvec)=G(\xvec+\Deltavec)$ for some $\Deltavec$ and all $\xvec$), it leads to a two-sample location problem, where we test the null hypothesis $H_0: \Deltavec={\bf 0}$ against the alternative $H_1: \Deltavec \neq {\bf 0}$. This is a well studied problem with wide spread applications. For instance, in clinical trials and other case control studies, this is often used to test the statistical significance of a treatment effect, where $F$ and $G$ correspond to the distribution of the measurement vector in the control group and the treatment group, respectively.
If $F$ and $G$ are assumed to be Gaussian, the $t$-statistic or the Hotelling's $T^2$ statistic (see e.g., Rao, 2001;  Anderson, 2003) are often used to perform the test. 
If we do not assume any parametric structure for $F$ and $G$, we need to use nonparametric methods. 
For instance, in the univariate case, rank based nonparametric tests like the Wilcoxon-Mann-Whitney (WMW) test and the Kolmogorov-Smirnov (KS) test (see e.g., Hollander and Wolfe, 1999) are commonly used. Several rank based methods 
are available for multivariate two-sample location problems as well (see e.g., Puri and Sen, 1971; Randles and Peters, 1990; Liu and Singh, 1993; Hettmansperger and Oja, 1994; Motonen and Oja, 1995; Choi and Marden, 1997; Hettsmanperger {\it et. al.}, 1998). A brief overview of these tests can be found in Oja and Randles (2004) and Oja (2010).

Most of these standard two-sample tests are mainly motivated by the ellipticity and unimodality of $F$ and $G$, and they yield satisfactory performance if $F$ and $G$ are nearly identical except for their locations. But, in practice, we often have situations, where $F$ and $G$, being  mixtures of several distributions, are multimodal in nature. Several authors have analyzed such mixture distributions and discussed about their applications in different fields of statistics (see e.g., Titterington, 1990; Gupta and Kabe, 1999; Chen et. al., 2004).
Suppose that we want to test the efficacy of a drug on a population. If this population happens to be a mixture of some ethnic groups, each of $F$ and $G$ may turn out to be a mixture of different sub-populations, where each sub-population corresponds to a particular ethnic group. Now, if the effect of the ethnicity on the study variables is higher than the treatment effect, direct use of the standard two-sample tests often gives misleading results. This problem becomes more evident if different ethnic groups have widely different proportion of representatives in the samples obtained from $F$ and $G$. 
In practice, we rarely get an ideal situation where $F$ and $G$ vary only due to the treatment effect. Always there are some hidden factors that can have influence on the study variables. In such cases, the use of standard two-sample location tests can be misleading, especially when the combined effect of these factors dominates the treatment effect. In this article, we investigate such cases and propose some methods
that can be used to modify the standard two-sample tests so that they can work successfully even in the presence of such hidden factors.

To demonstrate the importance of this study, let us consider a simple example. Suppose that we want to test the significance of a treatment effect on a population, which is an equal mixture of two sub-populations, and
the distributions of the measurement variable in these two sub-populations are $N(0,0.25)$ and $N(3, 0.25)$. Here $N(\mu, \sigma^2)$ denotes a normal (Gaussian) distribution with mean $\mu$ and variance $\sigma^2$. Assume that $p_1$ proportion of individuals from the first sub-population and $p_2$ proportion of individuals from the second sub-population are exposed to a treatment, and the rest belong to the control group. Also assume that the treatment
has the same effect on each of these sub-populations, and in each sub-population, it is supposed to increase the average value of the measurement variable by an amount $\delta$. So, while $F$ is a mixture of $N(0,0.25)$ and $N(3, 0.25)$ with mixing proportions $p_1$ and $p_2$, $G$ turns out to be a mixture of $N(\delta, 0.25)$ and $N(3+\delta, 0.25)$ with mixing proportions $1-p_1$ and $1-p_2$, respectively. Here we test  $H_0:\delta=0$ against $H_1:\delta>0$.

Let us first consider the case $\delta=0$. In this case, if $p_1$ is smaller than $p_2$ and the samples are drawn randomly, most of the sample observations from $F$ (and $G$, respectively) come from the sub-population having the lower (and the higher, respectively) value of the measurement variable. As a result, the direct use of a two-sample location test is expected to give a false alarm.
We carried out our experiment with $p_1=0.25$ and $p_2=0.75$, where 100 observations were taken from each of $F$ and $G$ to perform the test. The light and the dark grey curves in Figure 1(a) show the density functions corresponding to $F$ and $G$, respectively. Recall that in $t$-test, we assume the normality of the underlying distributions and test for the equality of their means. Figure 1(b) shows the estimated density functions of $F$ and $G$ under the normality assumption. It is quite clear from this figure that the mean of $G$ is much higher than that of $F$. So, the $t$ test rejected the null hypothesis $H_0: \delta=0$. We observed the same phenomenon for WMW and KS tests as well. Figure 1(c) shows the histogram of the pair-wise differences $\{d_{ij}=x_{2j}-x_{1i},~i=1,\ldots,n_1;~j=1,\ldots,n_2\}$ between the
observations from $F$ and $G$. Since most of the differences were positive (indicated using the dark grey color), the WMW test rejected $H_0$. Also, the dominance of the empirical version of $F$ (i.e., the empirical distribution function) over that of $G$ is quite evident in Figure 1(d). So, as expected, the KS test rejected the null hypothesis as well.

\begin{figure}[htp]
\centerline
{\includegraphics[height=1.60in,width=7.50in]
{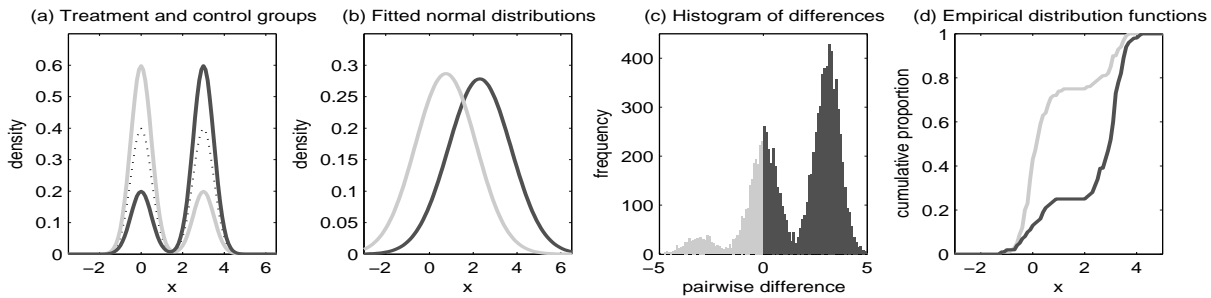}}
\vspace{-0.25in}
\caption{
{\small Performance of standard parametric and nonparametric tests ($\delta=0$).}}
\end{figure}

Now, consider the case $\delta =1$. Note that if $p_1$ is much larger than $p_2$, most of the observations from $G$ come from the sub-population having relatively lower values of the measurement variable. As a result, the usual two-sample tests often fail to detect the significance of the treatment effect. That is what we observed when we carried out our experiment with $p_1=0.75$ and $p_2=0.25$. Like before, we used 100 observations each from $F$ and $G$, and the results of our analysis are
presented in Figure 2. From this figure, it is quite clear that none of the standard parametric (t-test) and nonparametric (WMW and KS tests) tests rejected the null hypothesis.

\begin{figure}[htp]
\centerline
{\includegraphics[height=1.60in,width=7.50in]
{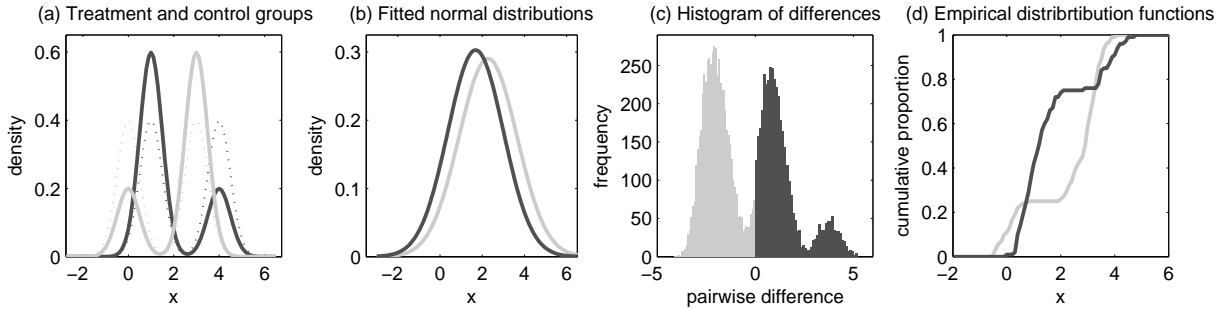}}
\vspace{-0.25in}
\caption{
{\small Performance of standard parametric and nonparametric tests ($\delta=1$).}}
\end{figure}

Not only for univariate problems, one can observe similar phenomenon for multivariate cases as well,
which we will see later.
In this article, we make an attempt to overcome these limitations of the standard methods in testing the significance of a treatment effect. 

\section{Description of the methodology}

Let ${\cal X}_1=\{\xvec_{11},\xvec_{12},\ldots,\xvec_{1n_1}\}$ and ${\cal X}_2=\{\xvec_{21},\xvec_{22},\ldots,\xvec_{2n_2}\}$ be independent realizations of the measurement vector $\Xvec$ obtained from  $F$ and $G$, respectively. In order to test
the significance of the treatment effect based on these two sets of independent observations ${\cal X}_1$ and ${\cal X}_2$,
we usually consider a location model (i.e., $F(\xvec)=G(\xvec+\Deltavec)$ for all $\xvec
\in R^d$) and use a standard parametric or nonparametric method to test  $H_0:\Deltavec ={\bf 0}$ against
$H_1:\Deltavec \neq {\bf 0}$ (in the univariate case, one can also consider one-sided alternatives). But as we have demonstrated in Section 1, these standard
tests often yield misleading results if the populations corresponding to $F$ and $G$ have several clusters or sub-populations, and the proportions of representatives from these sub-populations in ${\cal X}_1$ and ${\cal X}_2$ are widely
different. Suppose that $F$ and $G$ both have $k$ sub-populations ($k$ is unknown), and the distribution of the measurement vector $\Xvec$ in these sub-populations are denoted by $F_1,F_2,\ldots,F_k$ and $G_1,G_2,\ldots,G_k$, respectively. Here we consider a more general location model $F_r(\xvec)=G_r(\xvec+\Deltavec)$ for all $\xvec$ and $r=1,\ldots,k$, and carry out a test for $H_0:\Deltavec ={\bf 0}$ against $H_1:\Deltavec \neq {\bf 0}$. Note that here the mixing proportions in $F$ and $G$ may not be the same. 

\subsection{Elimination of sub-population effects}

Form our discussions in Section 1, it is quite clear that the presence of several sub-populations in the control and the treatment groups is the main source of the problem. If the number of sub-populations $k$ is known, and if we know which of these sub-populations each $\xvec_{ji}$ is coming from, this problem can be resolved easily. For instance, one can make suitable transformations of the observations
to merge these sub-populations before using any standard two-sample test. Let us assume that $F$ has
$k$ sub-populations, and $F_r$ is Gaussian with the mean $\muvec_r$ and the scatter $\sigmat_r$ ($r=1,2,\ldots,k$). Therefore, if $H_0$ holds, ${\cal X}_1 \cup {\cal X}_2$ will contain observations  from $k$
different clusters. Now, if we know that $\xvec$ is from $r$-th cluster ($r=1,2,\ldots,k$), the transformation $\xvec^{*}=\sigmat_r^{-1/2}(\xvec -\muvec_r)$ will give us an observation from the standard normal distribution. In
this way we can transform all $n_1$ observations from $F$ and $n_2$ observations from $G$. Under $H_0$, all of them
will follow the standard normal distribution. But if $H_0$ does not hold, the distributions of the transformed observations from $F$ and $G$ will differ. Note that unlike $F$ or $G$, the distributions of these transformed observations do not have further sub-populations. 
This transformation helps us to eliminate the effect of different clusters on the measurement variables.  Instead of normal, one can assume any other suitable parametric models for the sub-populations. However, as long as the sub-populations differ only in their locations and scale, the same transformation will work. For other parametric models, similarly one can find the appropriate transformations. If we do not assume any parametric structure for the sub-populations, we can use a transformation based on the spatial quantiles (see e.g., Chaudhuri, 1996; Koltchinskii, 1997) so that after the transformation, measurement vectors from different sub-populations have the same spatial quantiles. However, in practice, we do not know how many clusters there are, and which of the clusters an observation is coming from. So, in that case, one can use a suitable clustering algorithm to find the clusters in ${\cal X}_1 \cup {\cal X}_2$, and after finding the clusters, s/he can use the observations in each cluster to estimate the mean vectors and the scatter matrices, and hence the transformation function. We will discuss that in Sections 2.3 and 2.4.

\subsection{Test based on transformed observations}

After finding the clusters based on ${\cal X}_1$ and ${\cal X}_2$, if we make the transformation discussed above, the transformed variables in each group (control and treatment) can be as viewed as independent observations from a distribution with no sub-clusters. Therefore, one can use any standard two-sample test on
these transformed observations. Note that here we make the transformations based on the full set of observations ${\cal X}_1 \cup {\cal X}_2$. So,
in the univariate case, if we compute the WMW statistic or the KS statistic based on the transformed observations, because of the exchangeability
of the observations, the test statistic will still have the distribution-free property. Therefore, one can perform the WMW and the KS tests using the standard statistical tables. However, the multivariate non-parametric tests we considered in this article do not have the distribution free property. In such cases, one can either use the conditional test based on the permutation principle (see e.g., Puri and Sen, 1971; Hollander and Wolfe, 1999) or the test based on large sample distribution of the test statistic. In this article, we adopted the permutation method. The test performed
in this way removes the sub-population or cluster effect, and it usually yields better results, which we will see later. One should also notice that since the clustering is done based on all $n=n_1+n_2$ observations, class labels of the observations do not have any influence on the formation of the clusters. So, one does not need to run the clustering algorithm repeatedly for different permutations. The permutation principle can be used directly on the
transformed variables. This leads to a substantial saving in the computing time.

\subsection{Choices for the clustering algorithm and the number of clusters}

For the implementation of our method, one needs to estimate the number of sub-populations $k$ as well. We can run any suitable clustering algorithm (see e.g., Ripley, 1996; Duda {\it et. al.}, 2001; Hastie {\it et. al.}, 2009) for different numbers of sub-populations and choose the value of $k$ using an appropriate cluster validation index. In this article, we use the PAM (partitioning around medoids) algorithm (see e.g., Ripley, 1996; Hastie {\it et. al.}, 2001) for clustering. It is based on the $k$-medoids algorithm and preferred over the $k$-means algorithm because of its robustness against outliers. We use the PAM algorithm for different choices of $k$, and finally $k$ is chosen using the Dunn index (see Dunn, 1974). Note that Dunn index can be used to choose a value of $k$ when it is known that there are at least two clusters. So, if $k=2$ is selected, we use the gap statistic (see Tibshirani {\it et. al.}, 2001) to choose between $k=1$ and $k=2$. Of course, one can directly use the gap statistic to choose the value of $k$, but here we avoid that to reduce the computing cost.
In the presence of outliers in the data, often we get clusters consisting of one or a very few observations. Those observations, which belong to a cluster containing
less than $2\%$ of the observations, are treated as outliers, and we ignore them while determining the number of clusters. Jornsten (2004) used similar method for clustering and classification based on data depth. First we use this method separately on ${\cal X}_1$ and ${\cal X}_2$ to estimate the number of clusters in $F$ and $G$, respectively. Let these numbers be $k_1$ and $k_2$, respectively. Note that under the location model assumed at the beginning of this section, if not same, $k_1$ and $k_2$ are expected to be quite close. If they turn out to be same ($k_0$, say), we consider $k_0$ as the final value of $k$ and use PAM algorithm on ${\cal X}_1 \cup {\cal X}_2$ to find the final clusters. If $k_1$ and $k_2$ differ, we choose $k_0=\min\{k_1,k_2\}$. This often prevents us from false detection of clusters. Suppose that $F$ and $G$ both are unimodal and there is a statistically significant treatment effect. In such cases, if we
use any clustering algorithm on ${\cal X}_1 \cup {\cal X}_2$, we may find two clusters, one for each group. Now, if we use the transformation based on these two clusters, the treatment effect may get eliminated, and as a consequence, we may fail to detect the treatment effect even though it is statistically significant. But if we use the clustering algorithm on ${\cal X}_1$ and ${\cal X}_2$ separately, both $k_1$ and $k_2$ are likely to be $1$. As a result, $k_0=1$ is used, and the treatment effect is preserved.

In some cases, clustering based on ${\cal X}_1 \cup {\cal X}_2$ may lead to some clusters having almost all observations from one group. This gives an indication of a strong treatment effect. Such situations
may arise in some rare cases where each of the control and the treatment groups has two or more sub-populations (i.e., $k_0 \ge 2$) and the effect due to the sub-population difference is negligible compared to the treatment effect. In such cases, the above
transformation may have an adverse effect, and we do not recommend it. For instance, if the control group is a
mixture of $N(0,0.5)$ and $N(2,0.5)$ and the treatment group is a mixture of $N(10,0.5)$ and $N(12,0.5)$,
we may select $k_0=2$, and in that case, each of the two clusters on ${\cal X}_1 \cup {\cal X}_2$ will be formed by observations from one group only. However, in such cases, detection of treatment effect is rather easy, and one can directly use any standard two-sample test to get the desired result or even without a formal test, using clustering on ${\cal X}_1\cup{\cal X}_2$, one gets enough indication to reject $H_0$. Our proposed method is particularly helpful when the problem is more difficult and treatment effect is either
zero or so small that it gets dominated by the sub-population effect. So, we recommend to use our method when
each of the clusters obtained from ${\cal X}_1 \cup {\cal X}_2$ contains a certain proportion ($\beta$, say) of observations from each group. In this article, we used $\beta=0.1$, and during our data analysis, we faced this situation only once while analyzing the synthetic data in Section 5. In that case, we used the usual two-sample tests, and they rejected 
the null hypothesis.

\subsection{Model based clustering}

If we assume some parametric models for the sub-population distributions, the clustering algorithm can also be modified further. In that case, one can use the expectation-maximization (EM) algorithm (see e.g., Dempster, Larid and Rubin, 1977) to estimate the model parameters. After estimating these parameters, for any observation $\xvec$, we can estimate the posterior probabilities for different clusters, and the observation $\xvec$ can be assigned to the cluster having the largest estimated posterior. For estimating these posteriors, here we use equal priors for all $k_0$ sub-populations. It is non-informative and gives no preference to any of the sub-populations. Note that the EM algorithm needs the number of clusters $k_0$ and the initial values of the model parameters to be specified. For choosing $k_0$, here we use the method based on the Dunn index and the gap statistic as described before. We use the PAM algorithm with $k=k_0$ to find the initial clusters, and the initial estimates of the model parameters are computed from that. In the case of model based clustering, one can also run the EM algorithm for different numbers of clusters and select the value of $k_0$ using other model selection criteria.
However, to avoid extra computations involved in the repeated use of the EM algorithm, here we do not adopt these methods. 
In the case of Gaussian mixture models, some large sample tests have been proposed in the literature (see. e.g., Chen et. al., 2004; Chen and Li, 2009) to test the number of clusters, but they do not always guide us to choose $k_0$.
Throughout this article, we assume the sub-population distributions to be Gaussian and use the EM algorithm for
estimating the model parameters that lead to the final clustering.

\subsection{Results on the simulated example with mixture normal distributions}

Let us recall the simulated example discussed in Section 1, where $F$ was a mixture of $N(0,0.25)$ and $N(3, 0.25)$ with mixing proportions $p_1$ and $p_2$, and $G$ was a mixture of $N(\delta, 0.25)$ and $N(3+\delta, 0.25)$ with mixing proportions $1-p_1$ and $1-p_2$. First consider the case $p_1=0.25$, $p_2=0.75$ and $\delta=0$. In this case, the $t$-test, the WMW test and the KS test all rejected the true null hypothesis $H_0:\delta=0$. But when we used these tests on the transformed observations, all of them yielded the right decision. Figure 3 shows the performance of these tests when they were applied on the transformed observations. From this figure it is quite evident that there was no visible difference between the distributions of the transformed variables from the two groups. Generating different samples from $F$ and $G$, we repeated this experiment 100 times. In all these cases, the $t$-test, the WMW test and the KS test rejected the null hypothesis. But when they were used on transformed observations, they rejected $H_0$ in 5, 4 and 4 out of these 100 cases, respectively. Given that we fixed the nominal level of significance of these tests at 0.05, this result is quite encouraging.

\begin{figure}[htp]
\centerline
{\includegraphics[height=1.750in,width=7.250in]
{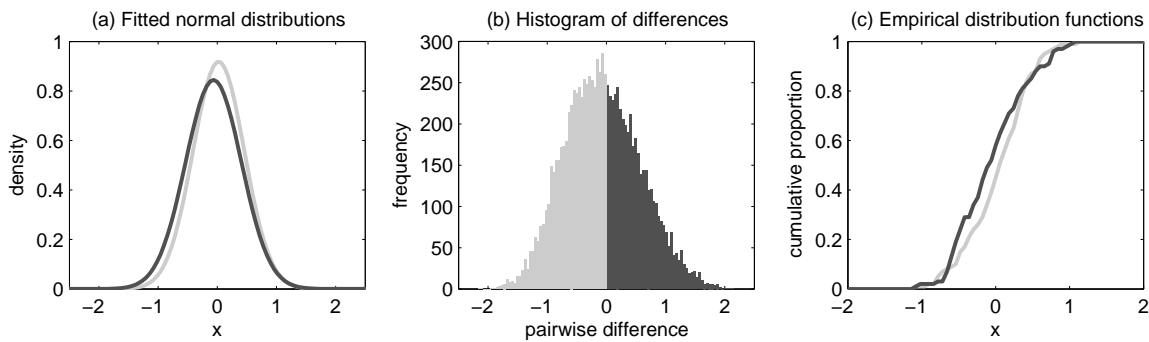}}
\vspace{-0.25in}
\caption{
{\small Performance of parametric and nonparametric tests after elimination of cluster effect ($\delta=0$).}}
\end{figure}

Next we consider the case $p_1=0.75$, $p_2=0.25$ and $\delta=1$. Recall that in this case, the $t$-test, the WMW test and the KS test all failed to detect the statistical significance of the treatment effect. But when we used these tests on the transformed observations, they led to the right decision. This is quite transparent from Figure 4. We repeated this experiment 100 times, and in all these cases, the transformation of the variable led to the rejection of the null hypothesis by all these methods. But when these standard tests were blindly used on the original observations, they had poor performance. The $t$-test and the WMW test failed to reject the null hypothesis even in a single occasion, while the KS test could reject it in 39 out of 100 cases.

\begin{figure}[htp]
\centerline
{\includegraphics[height=1.750in,width=7.250in]
{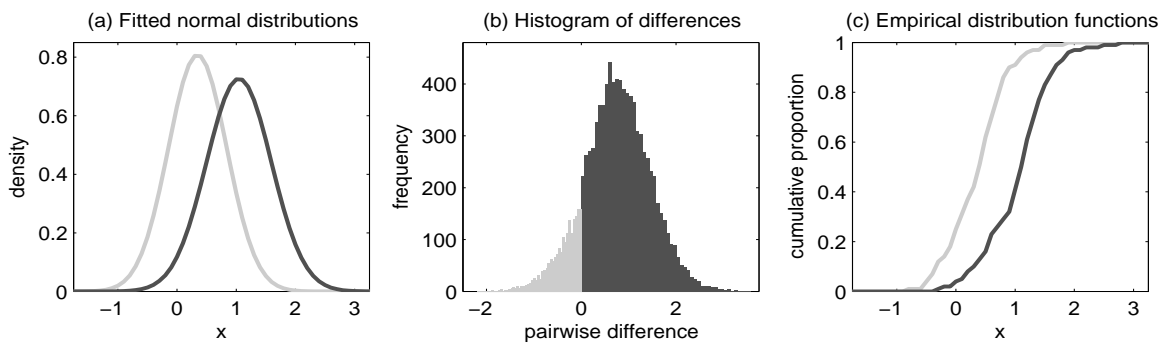}}
\vspace{-0.25in}
\caption{
{\small Performance of parametric and nonparametric tests after elimination of cluster effect ($\delta=1$).}}
\end{figure}

Note that if we assume that the sub-populations differ only in their location, after finding the clusters, one can also perform a two-way analysis of variance (ANOVA) considering the treatment and the cluster as the two factors. When we carried out this analysis, it yielded satisfactory performance. In the case of $\delta=0$ ($p_1=0.25, p_2=0.75$), it rejected the null hypothesis $H_0$ in 6 out 100 cases, but in the case of $\delta=1$ ($p_1=0.75, p_2=0.25$), it rejected $H_0$ in all 100 occasions.

\section{A modified method based on Bayesian model averaging}

The method described in the previous section works well when the clusters (sub-populations) are well separated. But if there are several overlapping clusters, depending on the cluster labels assigned to the observations in the overlapping region, it may lead to different inferences. Suppose that we have a sample from a univariate population with two overlapping sub-populations. Now, if an observation in the overlapping region is assumed to come from the left cluster
(the cluster formed by relatively lower values of the variable), after the transformation, it will lie at the right tail of the distribution of the transformed variable, but if we consider it as an observation from the other cluster, after the transformation, it will become an observation at the left tail. Clearly, the value of the parametric and the non-parametric test statistic will depend on that. So, if we have a reasonably large number of observations in the overlapping region, depending on the cluster levels assigned to them, the method described in the previous section can lead to diametrically opposite inferences. In order to overcome this limitation, here we modify our algorithm using the Bayesian model averaging technique (see e.g., Hoeting {\it et. al.}, 1996; Claeskens and Hjort, 2008).

Let ${\cal X}_1$ and ${\cal X}_2$ be the data clouds from the control and the treatment groups (i.e., $F$ and $G$), respectively, and let us define a random variable $Z$, which takes the value $1$ when $H_0$ is rejected, and $0$ if it is accepted. In usual two-sample tests, we do not search for the sub-populations and compute the test statistic $T({\cal X})$ and the corresponding test function $\phi({\cal X})=P(Z=1\mid {\cal X})$
based on ${\cal X}=({\cal X}_1,{\cal X}_2)$ only. However, in our method, we take the cluster labels of the observations into consideration. Let ${\cal C}_1=(c_{11},\ldots,c_{1n_1})$ and ${\cal C}_2=(c_{21},\ldots,c_{2n_2})$ be the cluster labels corresponding to ${\cal X}_1$ and ${\cal X}_2$, respectively. Now based on this particular assignment of clusters, we can make the transformation of the variables as discussed above, and compute the test statistic and the associated test function. But, in addition to ${\cal X}$, this test function depends on the choice of ${\cal C}=({\cal C}_1,{\cal C}_2)$, and instead of computing $P(Z=1\mid {\cal X})$, we actually compute the conditional probability $P(Z=1\mid {\cal X}, {\cal C})$. Clearly the resulting test
statistic depends on the choice of ${\cal C}$, and depending on the nature of the problem and the separability
among the clusters, the final inference can be sensitive on this choice.
In order to overcome this model uncertainty, we can use the Bayesian model averaging technique that aggregates the
results for the all possible choices of ${\cal C}$. Similar approaches based on Bayesian model averaging have also been used for
supervised and semi-supervised classification based on kernels and nearest neighbors (see e.g., Holmes and Adams, 2002; Mukhopadhyay and Ghosh, 2011). From the results of elementary probability theory, we have
\[  \phi({\cal X})=P(Z=1\mid {\cal X}) = \sum_{{\cal C} \in \{1,2,\ldots,k_0\}^n} P(Z=1\mid {\cal X}, {\cal C}) ~P({\cal C} \mid {\cal X}),
\]
where $P({\cal C}\mid{\cal X})$ is the conditional probability distribution of ${\cal C}$ given ${\cal X}$. Now, from the Bayes' theorem, we know that $P({\cal C} \mid {\cal X}) = \pi({\cal C}) f({\cal X} \mid {\cal C})/ \sum_{\cal C} \pi({\cal C}) f({\cal X} \mid {\cal C})$,
where $\pi ({\cal C})$ is the prior distribution of ${\cal C}$, and $f({\cal X} \mid {\cal C})$ is the conditional density of ${\cal X}$ given ${\cal C}$. Throughout this article, we consider a uniform prior distribution for  ${\cal C}$. Note that this prior is non-informative and it gives no preference to any of the $k_0$ clusters.
For this choice of prior, we have $P({\cal C} \mid {\cal X}) = f({\cal X} \mid {\cal C})/ \sum_{\cal C} f({\cal X} \mid {\cal C})$. If $n^{k_0}$ is not large, we can compute the test function $P(Z=1\mid {\cal X}, {\cal C})$ for all possible $n^{k_0}$ choices of ${\cal C}$, and using the formula given above, $\phi({\cal X})$ can be obtained as an weighted average of those conditional test functions. If $n^{k_0}$ is large, it may not be computationally feasible to
consider all possible choices of ${\cal C}$, In that case, we can generate sufficiently large number of observations ${\cal C}^{(1)},\ldots,{\cal C}^{(M)}$ from the posterior distribution $P({\cal C} \mid {\cal X})$, and approximate $\phi({\cal X})$ by its sample analog, which is given by \[ \phi({\cal X}) \simeq \frac{1}{M} \sum_{m=1}^{M} P(Z=1\mid {\cal X}, {\cal C}^{(m)}).\] 

In practice, in order to compute the weighted average or to generate from $P({\cal C} \mid {\cal X})$, we need to estimate the unknown quantity $f({\cal X} \mid {\cal C})$ for different choices of ${\cal C}$. If $f_1,\ldots,f_{k_0}$
denotes the sub-population distributions, $f({\cal X} \mid {\cal C})$ can be approximated by the psudo-likelihood
function $\prod_{i=1}^{n_1} \prod_{j=1}^{n_2} {\hat f}_{c_{1i}}(\xvec_{1i}) {\hat f}_{c_{2j}}(\xvec_{2j})$, where
${\hat f}_r$ is the estimate of the density function $f_r$ for $r=1,2,\ldots,k_0$. If we assume a parametric
model $f_j(\cdot)=\psi_j(\cdot,\theta_j)$ for $f_j$, where $\psi_j (\cdot)$ is a known function and $\thetavec_j$
is a unknown parameter (scalar or vector valued), we find an estimate ${\widehat {\thetavec}}_j$ from the data to compute ${\hat f}_j(\cdot)=\psi_j(\cdot,{\widehat {\thetavec}}_j)$. Otherwise, a nonparametric estimate can be used.
In this article, we assume these sub-population distributions to be Gaussian, and the mean vectors and dispersion matrices of these distributions are estimated using the EM algorithm. This Bayesian model averaging technique makes the final decision free from the choice of ${\cal C}$, and it usually improves the performance of the resulting test procedure, which we will see in the subsequent sections.

Note that if the clusters are well separated, we have $P({\cal C} \mid {\cal X})\simeq 1$ for one particular
choice (correct choice) of ${\cal C}$, while for other choices of ${\cal C}$, it becomes close to zero. In that case,
our modified method matches with the method described in Section 2. So, it can be viewed as a generalization of the method described earlier. In this article, we used both of these methods for
our data analysis. Henceforth, the earlier method will be referred to as Method-1, and the method based on Bayesian model averaging will be referred to as Method-2. One should also notice that if there is only one cluster in the data, we do not need to make any transformation at all. In those cases, our tests based on the WMW statistic and the KS statistic coincide with their usual versions, and those based on Hotelling's $T^2$ (or $t$-statistic in the univariate case), coordinate-wise ranks and spatial ranks differ from their usual versions only in the method of computing the cut-off values. So, in those cases, our proposed methods are expected to perform like the usual two-sample tests. In Section 5, our analysis of Iris data will make it more clear.


\section{Results from the analysis of simulated datasets}

Using some simulated examples, in the earlier sections, we have already demonstrated the importance of our proposed methods in the case of univariate one sided alternatives. In this section, we use some univariate and bivariate simulated examples to evaluate its performance for two-sided alternatives. In the univariate set up, we use the $t$-test, the WMW test and the KS test as before. In the multivariate case, we use a parametric test based on the Hotelling $T^2$ statistic (see e.g., Rao, 2001; Anderson, 2003) and two non-parametric tests based on
coordinate-wise ranks (e.g., Puri and Sen, 1971) and spatial ranks (see e.g., Motonen and Oja, 1995; Choi and Marden, 1997).

Let us begin with an example involving mixture of univariate normal distributions, where $F$ is a mixture of  $N(0,0.25)$, $N(\mu,0.25)$ and $N(2\mu,0.25)$, and $G$ is a mixture of $N(\delta,0.25)$, $N(\mu+\delta,0.25)$ and $N(2\mu+\delta,0.25)$. The proportion of mixing in $F$ and $G$ were taken as $0.4, 0.3, 0.3$ and $0.6, 0.2, 0.2$, respectively. We used samples of size 100 from each group to test $H_0:\delta=0$ against $H_1:\delta \neq 0$, and repeated the experiment 100 times. We considered two values of $\mu$ (2 and 3) and five values of $\delta$ (-0.5, -0.25, 0, 0.25 and 0.5), and the results of different tests are reported in Table 1. From this table, one can easily assess the importance of the transformation to eliminate cluster effect. In the case of $\delta=0$, the usual parametric and non-parametric tests failed to maintain the nominal level of 0.05, but Method-1 and Method-2 had good level properties. The usual tests were biased towards negative values of $\delta$. As a result, they rejected $H_0$ too often when we considered $\delta=-0.5$ and $\delta=-0.25$. But, for positive values of $\delta$, they had poor power properties. In this example, Method-1 had reasonably good performance, but Method-2 performed even better. Note that
in cases with $\mu=3$, due to higher difference among the sub-populations, clustering becomes
much easier. So, as expected, the proposed methods yielded superior performance in those cases.



{\small
\begin{table}[h]
\caption{Proportion of times $H_0:\delta=0$ is rejected in the case of univariate normal mixtures}
\centering
{\small{
\begin{tabular}{|c|ccccc|ccccc|ccccc|}
\hline
& \multicolumn{5}{@{\hspace{0.05in}}c@{\hspace{0.05in}}|}{$t$-test} & \multicolumn{5}{@{\hspace{0.05in}}c@{\hspace{0.05in}}|}{WMW test} & \multicolumn{5}{@{\hspace{0.05in}}c@{\hspace{0.05in}}|}{KS test}\\ \hline
$\delta$ & -0.5 & -0.25  & 0 & 0.25 & 0.5   & -0.5 & -0.25  & 0 & 0.25 & 0.5 & -0.5 & -0.25  & 0 & 0.25 & 0.5 \\ \hline
\multicolumn{16}{|c|}{~~~~~~~~~~$\mu$=2}\\ \hline
Usual test &.99 &.93 &.78 &.25 &.07 &1.0 &.96 & .79  &.17 &.08 &1.0 &.93 &.77 &.49 &.44 \\ \hline
Method-1   &.87 &.74 &.04  &.40  &.67 &.88  &.70 &.07 &.41 &.68 &.82 &.59 &.05 &.34 &.70\\ \hline
Method-2   &.90 &.71 &.06 &.41  &.73 &.91 &.69 & .06 &.45 &.78 &.86 &.56 &.05 &.39 &.79 \\ \hline
\multicolumn{16}{|c|}{~~~~~~~~~~$\mu$=3}\\ \hline
Usual test &.97 &.92 &.65 &.27 &.07 &1.0 &.97 & .64& .18& .06&1.0 &.92 & .63& .49& .40\\ \hline
Method-1 &.95 &.84 & .04 & .50 & .91&.97  &.84 &  .03& .54& .96&1.0 &.66 &.03 &.51 &.94\\ \hline
Method-2 &.98 &.81 &.06 &.61 & .99&.99 &.83 &.05 &.61 &.98 &1.0 &.64 &.04 &.52 &.97 \\ \hline

\end{tabular}}}
\end{table}
}


Next we consider some examples involving mixtures of bivariate normal distributions. Here $F$ is a mixture of $N_2({\bf 0},\sigmat)$, $N_2(\muvec,\sigmat)$ and $N_2(2\muvec,\sigmat)$ and $G$ is a mixture of $N_2({\bf 0}+\Delta,\sigmat)$, $N_2(\muvec+\Delta,\sigmat)$ and $N_2(2\muvec+\Delta,\sigmat)$, where ${\bf 0}=(0,0)^{'}$, $\muvec=(3,3)^{'}$, and $\sigmat$ is the $2 \times2$ symmetric matrix with two diagonal entries equal to 1 and the off-diagonal entry -0.5. We considered $\Delta$ of the form $\Delta=(\delta,\delta)$ and studied the performance of different tests for five
different choices of $\delta$ as used in Table 1. We used the same mixing proportions for $F$ and $G$
as in the univariate case, and each experiment was repeated 100 times as before. Here also we observed the same phenomenon as in the univariate case. While direct applications of the usual two-sample tests failed to maintain the nominal level of 0.05 and led to poor power properties for positive values of $\delta$, the tests based on Method-1 and Method-2 yielded better performance both in terms of level and power properties. In Table-2, one can also observe that the tests based on our proposed methods had higher power for $\delta=-0.5$ and $-0.25$ as compared to that for the corresponding positive values of $\delta$. The reason for such asymmetry becomes clear from Figure 5, which shows the scatter plots of the observations in two groups (indicated using light grey and dark grey dots) in cases of $\delta=0.5$ and $\delta=-0.5$. In Figure 5(a), the three clusters are not that evident as they are in Figure 5(b). We observed this several times over the 100 simulations we carried out. Since clustering was much easier in the latter case, our proposed methods had relatively better performance for negative values of $\delta$.

\begin{figure}[htp]
\centerline
{\includegraphics[height=2.50in,width=7.0in]
{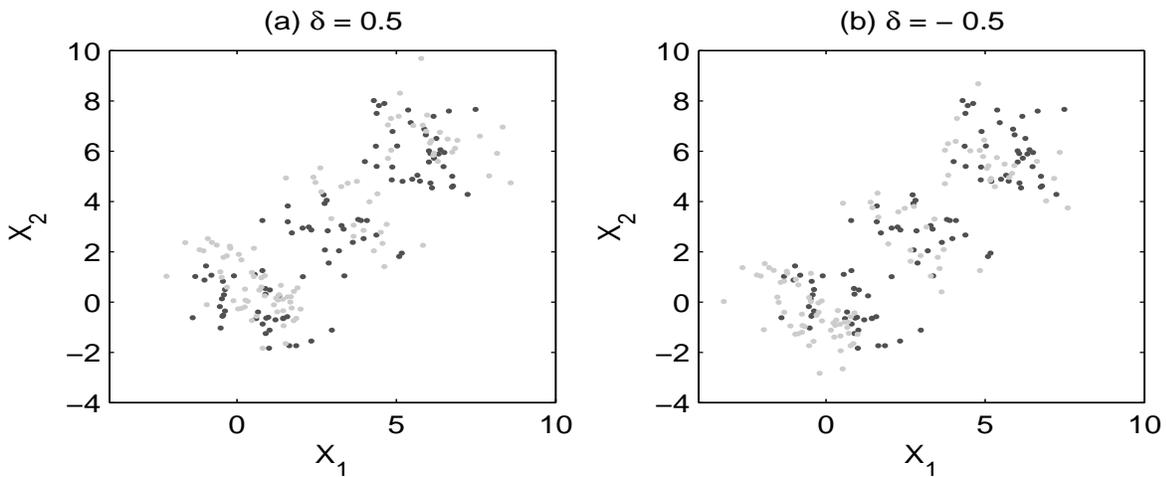}}
\vspace{-0.25in}
\caption{
{\small Scatter plot of observations from the two bivariate mixture normal distributions.}}
\end{figure}

{\small
\begin{table}[h]
\caption{Proportion of times $H_0:\delta=0$ is rejected in the case of bivariate normal mixtures}
\centering
{\small{
\begin{tabular}{|c|ccccc|ccccc|ccccc|}
\hline
& \multicolumn{5}{c|}{Hotelling $T^2$ test} & \multicolumn{5}{c|}{Coordinate-wise  rank test} & \multicolumn{5}{|c|}{Spatial rank test}\\ \hline
$\delta$ & -0.5 & -0.25  & 0 & 0.25 & 0.5   & -0.5 & -0.25  & 0 & 0.25 & 0.5 & -0.5 & -0.25  & 0 & 0.25 & 0.5 \\ \hline
Usual test& .92 & .82 & .61 &  .36& .15  & .96 &.84  &.63  &.37  &.17  &.96  &.86  &.63  &.35  &.16 \\ \hline
Method-1& .96 & .69 & .04 & .69 &.85 &.97  &.69  &.07  &.66  &.85& .98  & .67 & .04 & .69 & .87\\ \hline
Method-2& .99 & .70 &.05  &.70  &.92  &1.0  &.68  & .05 & .67 &.92  &1.0  &.68  &.06  &.69  &.93 \\ \hline
\end{tabular}}}
\end{table}
}

Keeping all location and scatter parameters unchanged, we repeated the same experiment with mixtures of Cauchy distributions. Because of the heavy tails of the
Cauchy distributions, we did not always get three distinct clusters. As a result, the proposed method could only have marginal improvements over the standard
methods. For instance, in the case of $\delta=0.5$, while direct application of the coordinate-wise rank test and the spatial rank test both led to the rejection of $H_0$ in 7 out of 100 cases, using Method-1 (Method-2),
we could reject it in 11 and 12 (16 and 18) occasions, respectively. So, for mixtures of bivariate Cauchy distributions, we chose $\muvec=(5,5)$ and carried out experiment for five choices of $\delta$ (-1, 0.5, 0, 0.5 and 1). The results in Table 3 show that even in this case, the proposed methods, especially Method-2 had better level and power properties.

{\small
\begin{table}[h]
\caption{Proportion of times $H_0:\delta=0$ is rejected in the case of bivariate Cauchy mixtures}
\centering
{\small{
\begin{tabular}{|c|ccccc|ccccc|ccccc|}
\hline
& \multicolumn{5}{c|}{Hotelling $T^2$ test} & \multicolumn{5}{c|}{Coordinate-wise rank test} & \multicolumn{5}{|c|}{Spatial rank test}\\ \hline
$\delta$ & -1.0 & -0.5  & 0.0 & 0.5 & 1.0  & -1.0 & -0.5  & 0 & 0.5 & 1.0 & -1.0 & -0.5  & 0 & 0.5 & 1.0 \\ \hline
Usual test& .52 & .48 & .31 & .16 & .06 & .98 & .93 & .60 & .20 & .07 & .98 & .93 & .58 & .16 &.05\\ \hline
Method-1& .14 & .04 & .03 & .04 & .12  &.55 & .50 & .02 & .31 & .50 & .73 & .62 & .05 & .35 & .49\\ \hline
Method-2 & .18 & .05 & .03 & .04 & .13 & .64 & .57 & .04 & .36 & .59 & .82 & .64 & .06 & .43 & .60\\ \hline
\end{tabular}}}
\end{table}
}

So, far we have considered the cases where $\muvec$ and $\Delta$ are in the same (or opposite) direction, and the mixing proportions in the control and the treatment groups are different. Now we consider the situation, where the
two groups have the same mixing proportion, and $\muvec=(3,3)$ and
$\Delta=(0,\delta)$ are not in the same direction. We consider the same bivariate normal sub-populations for $F$, but
the mixing proportions for $N_2({\bf 0},\sigmat)$, $N_2(\muvec,\sigmat)$ and $N_2(2\muvec,\sigmat)$ are taken as .3 .4 and .3, respectively.
We used the same mixing proportions for $N_2(\Delta,\sigmat)$, $N_2(\muvec+\Delta,\sigmat)$ and $N_2(2\muvec+\Delta,\sigmat)$ in $G$ as well. Here the powers of different tests do not depend on the sign of $\delta$. So, we report the results only for five non-negative values (0, 0.25, 0.5, 0.75 and 1) of $\delta$.
Unlike previous cases, here the usual parametric and nonparametric tests maintained their levels (see Table 4), but from Table 4, it is quite transparent that the proposed methods had better power properties.

{\small
\begin{table}[h]
\caption{Proportion of times the null hypothesis is rejected in the case of bivariate normal mixtures, where the alternative
suggests a shift in the direction of $y$-axis}
\centering
{\small{
\begin{tabular}{|c|ccccc|ccccc|ccccc|}
\hline
& \multicolumn{5}{c|}{Hotelling $T^2$ test} & \multicolumn{5}{c|}{Coordinate-wise rank test} & \multicolumn{5}{|c|}{Spatial rank test}\\ \hline
$\delta$ & 0 & 0.25 & 0.5 & 0.75 & 1 & 0 & 0.25 & 0.5 & 0.75 & 1 & 0 & 0.25 & 0.5 & 0.75 & 1 \\ \hline
Usual test & .05 &.21 &.45 &.80 &.96 &.07 &.20 &.45 &.79 &.94 & .06 &.20 &.48 &.82 & .96\\
Method-1 & .05 &.27 &.71 &.95 &.99 &.05 &.25 &.72 &.92 &.98 & .05 &.25 &.73 &.93 & .98\\
Method-2   & .05 &.28 &.73 &.97 &1.0 &.06 &.25 &.73 &.95 &.99 & .05 &.26 &.74 &.97 &1.0 \\ \hline
\end{tabular}}}

\end{table}
}

Next we consider a case, where the centers of the sub-population distributions do not lie on the same straight line, and these distributions differ also in the scatter. We chose $F$ to be a mixture of $N(0,0,1,0.5,0)$, $N(3,3,0.5,1,0)$ and $N(6,0,1,0.5,0)$ with mixing proportions 0.3, 0.4 and 0.3, respectively. Here $N(\mu_1,\mu_2,\sigma_1^2,\sigma_2^2,\rho)$ denotes a bivariate normal distribution
with the location vector $(\mu_1,\mu_2)^{'}$, marginal variances $\sigma_1^2$ and $\sigma_2^2$, and the correlation coefficient $\rho$. Here also, we consider the same mixing proportion (0.3, 0.4 and 0.3) for $G$, where $G(\xvec)=F(\xvec+\Delta)$ for all $\xvec$ and $\Delta=(0,\delta)$. We carried out this experiment for $\delta=-0.5,-0.25,0,0.25$ and $0.5$, and the results are reported in Table 5. The superiority of the proposed methods is quite evident from this table. Also, as it is expected, we got nearly the same result for positive and negative value of $\delta$.

{\small
\begin{table}[h]
\caption{Proportion of times the null hypothesis is rejected in the case of bivariate normal mixtures, where the sub-populations differ in
locations and scales}
\centering
{\small{
\begin{tabular}{|c|ccccc|ccccc|ccccc|}
\hline
& \multicolumn{5}{c|}{Hotelling $T^2$ test} & \multicolumn{5}{c|}{Coordinate-wise rank test} & \multicolumn{5}{|c|}{Spatial rank test}\\ \hline
$\delta$ & -0.5 & -0.25 & 0 & 0.25 & 0.5  & -0.5 & -0.25  & 0 & 0.25 & 0.5 & -0.5 & -0.25  & 0 & 0.25 & 0.5 \\ \hline
Usual test&.54 &.13 &.06& .14& .51 &.57 &.14 &.05 &.15 &.56 &.58 &.16 &.06 &.16 &.54\\ \hline
Method-1&.94 &.39& .07& .42 &.92 &.91 &.38 &.07 &.41 &.93 &.92 &.39 &.07 &.43 &.94 \\ \hline
Mehtod-2&.93 &.41& .07& .42 &.91 &.92 &.39 &.06 &.40 &.92 &.93 &.42 &.07 &.44 &.92 \\ \hline
\end{tabular}}}
\end{table}
}

\section{Results from the analysis of some benchmark data sets}

In this section, we analyze three benchmark data sets for further evaluation of the proposed methods. These data sets, namely Abalone data, Synthetic data and Iris data, and their descriptions are available at the UCI Machine Learning Repository ({\small http://www.archive.ics.uci.edu/ml/datasets.html}).

\subsection{Abalone data}

In this data set, though there are observations from 28 classes, for our analysis, we considered the two largest classes (class 9 and class 10) only. This data set contains information on seven continuous variables, but most of these variables are highly correlated, and that makes the estimated dispersion matrix nearly singular in most of the cases. That is why instead of
considering all these variables, we used the first principle component as our measurement variable. This principal component explained 97.5\% of the total variation. In addition to this, there was another variable that indicated whether the abalone was a male or a female or an infant. We did not use it for our analysis and considered it as a hidden factor.

In order to check the level properties of different tests, first we randomly chose two sub-samples each of size 100 from the observations in class-9. The first sub-sample was formed by taking 20 males, 20 females and 60 infants, while the second sub-sample consisted of 25 males, 50 females and 25 infants. These two sub-samples were used as observations from two groups to test $H_0:\Delta=0$ against $H_1:\Delta>0$. This experiment was carried out 200 times and the results are reported in Table 6. In this case, the usual tests based on the $t$-statistic, the WMW statistic and the KS all rejected the null hypothesis in more than 10\% of the cases. But, Method-1 and Method-2, especially, the later one, showed better level properties.

\begin{figure}[htp]
\centerline
{\includegraphics[height=3.750in,width=7.50in]
{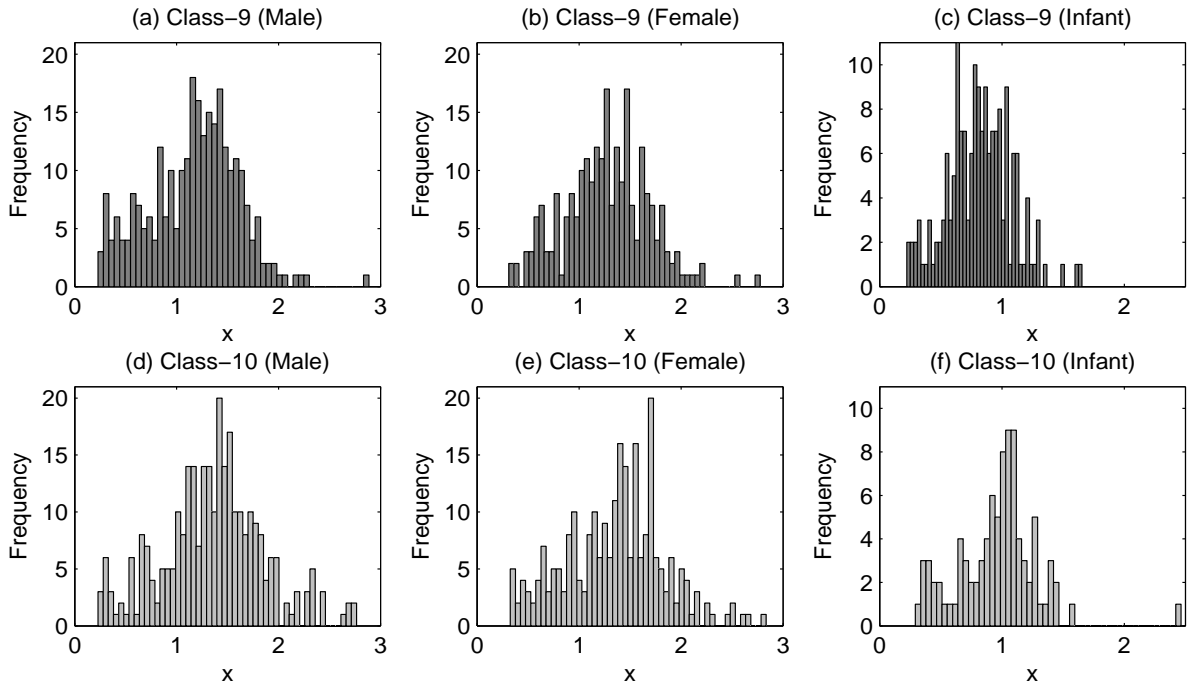}}
\vspace{-0.25in}
\caption{
{\small Histogram of the first principal component in Abalone data set.}}
\end{figure}

Next, we use the observations from both class-9 and class-10 to investigate the power properties of different tests. Figure 6 shows the histogram of the distribution of this first principal component in the three sub-populations (male, female and infant) of these two classes. In this figure, one can notice that in all the three cases (males, females and infants), the distribution in class-10 has a marginal shift towards the positive direction of the $X$-axis. So, naturally, one would be interested in testing $H_0:\Delta=0$ against $H_1:\Delta>0$. When we performed a two-sample test using all males from these two classes, the $t$-test, the WMW test and the KS test, all rejected $H_0$. We observed the same phenomenon for the females and the infants as well. So, ideally one would expect the null hypothesis to be rejected when observations from all these sub-populations are considered simultaneously. However, if we use the whole data set for testing, any test will either accept or reject $H_0$. Based on that single experiment, it will be difficult to compare among different test procedures. So, instead of using all observations, here we perform these tests based on randomly chosen subsets of size 100 from each class, and repeated the experiment 200 times.
In case of class-9, we chose observations from the three clusters with equal probability, but for class-10, these probabilities were 0.2, 0.2 and 0.6, respectively, for males, females and infants. In the presence of these sub-populations, direct application of the standard two-sample tests yielded poor power (see Table 6). Note that in this example, three sub-populations were highly overlapped. So, it was very difficult to find out these three clusters distinctly. In fact in many cases, the number of clusters was over estimated and in a few cases, it was under estimated too. But instead of that, the proposed methods performed better, and they rejected $H_0$  more often than the standard methods.

\subsection{Synthetic data}

In synthetic data set, each class is a mixture of two bivariate normal distributions differing only in their location. A scatter plot of this data set is given in Figure 5, where the dots ($\cdot$) and the crosses ($\times$) represent the observations from the two classes. Here we consider these two classes as the control and the treatment groups. From this figure, it is quite clear that for each sub-population, the shift is nearly in the same direction.
So, one should expect $H_0$ to be rejected. We randomly chose 30 observations from each group to
perform the test. In this example, the standard two-sample tests worked well, and they rejected $H_0$ in all occasions. Since the clusters were well separated, our clustering method rightly detected the
two sub-populations in each group in 197 out of 200 cases, while the number of clusters $k_0$ was over-estimated in other three cases.
So, as expected, the proposed methods performed well. Only in one occasion, we obtained a cluster containing less than 10\% observations from one group. In that case,
we used the usual two-sample tests and they rejected $H_0$.
Because of the multi-modal nature of the data, in this example, these proposed methods were expected to outperformed the traditional methods, but that did not happen. Note that
here the treatment effect is along the $Y$-axis, but in each group the sub-population effect is in an orthogonal direction along the $X$-axis. So, unlike the previous cases, this sub-population effect did not have any effect on the final outcome when the usual two-sample tests were used.

\begin{figure}[htp]
\centerline
{\includegraphics[height=2.50in,width=3.50in]
{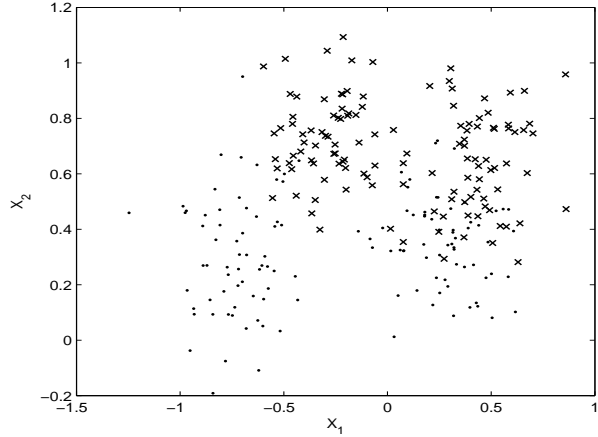}}
\vspace{-0.25in}
\caption{
{\small Scatter plot of Synthetic data.}}
\end{figure}

To investigate the level properties of different tests, we randomly generated two independent set of 30 observations from the class indicated using dots and used them as observations from control and treatment groups. We repeated this experiment 200 times, and all testing procedures rejected $H_0$
 in nearly 5\% of the cases. But, the result was quite different when the samples were generated in a different way. We randomly chose 30 observations from each of the two clusters. While 20 observations from the left cluster and 10 from the right cluster were used as observations from $F$, the rest 10 from the left and 20 from the right clusters were used as observations from $G$. In this case, as expected, the usual tests failed to maintain their levels  (see Table 6). The Hotelling $T^2$ test rejected $H_0$ in all 200 occasions. The coordinate-wise rank test and spatial rank test also rejected $H_0$ in more than 150 cases. However, our proposed methods had good level properties even in this situation. In all cases, the proportion of rejection was close to the nominal level of 0.05.

\subsection{Iris data}

Finally, we consider the Fisher's iris data. This data set contains four-dimensional observations from three different classes. For our analysis, we used observations from two classes: iris versicolor and iris verginica.
So far, we have carried out our analysis with data sets having
two or more hidden clusters. In this data set, the distributions corresponding to each of these two classes are known to be unimodal, and they have no hidden clusters. It is also well known that these two distributions differ in their location. So, the standard two-sample tests are expected to perform well. We considered this data set to see how the proposed methods perform in such cases. We randomly chose 25 observations from each class to constitute
the sample and perform two-sample tests based on these 50 observations. This experiment was repeated 200 times as before. In all these cases, the standard two-sample tests rejected the null hypothesis. When we used the clustering method, in all these 200 cases, only one cluster was selected. As a result, the proposed methods could match the
performance of the traditional ones. In order to investigate the level properties of different tests, we partitioned the observations in the versicolor class into two groups of size 25, which were used as observations from $F$ and $G$. This random partitioning was done 200 times as before, and for all these tests, the proportion of rejecting $H_0$ was close to the nominal level of 0.05.

{\small
\begin{table}[h]
\caption{Levels and powers of different tests in benchmark data sets.}
\footnotesize{
\centering
\begin{tabular}{|c c|c|c|c|c|c|c|c|c|c|c|} \hline
     &  & \multicolumn{3}{c|}{t-test} & \multicolumn{3}{c|}{WMW test} & \multicolumn{3}{c|}{KS test} \\ \cline{3-11}
     &  & Usual &  Meth. 1 &  Meth. 2 & Usual & Meth. 1&  Meth. 2  & Usual &  Meth. 1 &  Meth. 2  \\ \hline
Abalone & Level  &0.120 & 0.095 &0.060 &0.135 &0.080  &0.055 &0.155 &0.085  &0.070\\
         & Power &0.025 &0.235 &0.280 &0.050 & 0.245 &0.255 &0.065 & 0.370 &0.345\\ \hline
     &  & \multicolumn{3}{c|}{Hotelling $T^2$ test} & \multicolumn{3}{c|}{Coordinate-wise rank test} & \multicolumn{3}{c|}{Spatial rank test} \\ \cline{3-11}
     &  & Usual &  Meth. 1  & Meth. 2  & Usual &  Meth. 1  &  Meth. 2  & Usual &  Meth. 1  &  Meth. 2  \\ \hline
Synthetic & Level &1.000 & 0.045 &0.060 &0.770 & 0.030 &0.040 &0.755 & 0.040 &0.055 \\
         & Power  &1.000 & 0.995 &1.000 &1.000 &0.995  & 1.000  &1.000 & 0.990 &1.000\\ \hline
Iris     & Level  &0.045 & 0.045 &0.045 &0.060 & 0.060 &0.060 &0.055 &0.050  &0.050 \\
         & Power  &1.000 & 1.000 &1.000 &1.000 & 1.000 &1.000 &1.000 &1.000  &1.000\\
\hline
\end{tabular}}
\end{table}
}

\section{Concluding remarks}

This article shows the limitations of the usual two-sample parametric and nonparametric methods in testing the significance of a treatment effect. Direct use of these two-sample tests often yield misleading inferences,
especially when the underlying population has several hidden sub-populations, and the impact
of the sub-population difference on the measurement vector dominates the treatment effect. This problem becomes
even more evident if the control and the treatment groups have widely different proportions of representatives
from these sub-populations. The methods we proposed in this article take care of these problems. Using a
clustering algorithm, they detect the hidden clusters (sub-populations) in these groups and then
eliminate the cluster effect using a suitable transformation. So, in the presence of hidden sub-populations, while direct use of the standard
two-sample tests leads to unsatisfactory performance, our proposed methods perform better and yield reliable results. Moreover, when there are no hidden clusters,
these methods can perform as good as the standard two-sample tests. Using several simulated and benchmark data sets,
in this article, we have amply demonstrated these two important features of the proposed methods.

In this article, after estimating $k_0$, we used the EM algorithm to fit a mixture of $k_0$ Gaussian models to the data. The estimates of density functions corresponding to different clusters and hence the posterior probabilities of different clusters were computed from that. These estimated posteriors were used by the Bayesian model averaging method to take care of the uncertainty involved in the formation of $k_0$ clusters.
However, instead of working with a fixed $k_0$, one can also consider the results for different choices of $k_0$ and judiciously aggregate them. For instance, as an alternative, one can use Bayesian nonparametric density estimation using Gaussian mixtures with Dirichlet process prior. In that case, the number of clusters is also treated as random, and along with the parameters of sub-population distributions, it follows a joint prior distribution $D(\alpha, G_0)$, which is a Dirichlet process defined by a positive scalar $\alpha$ and a specified base distribution $G_0(\cdot)$ (see e.g, Escober and West, 1995). However, to keep our methods simple, here we did not consider them. Note that in comparison to Dirichlet modeling, these proposed methods are far more amenable for computation and they do not depend on a number of hyper-parameters.

\thebibliography{xx} 


\item Anderson, T. W. (2003) {\em An Introduction to Multivariate Statistical Analysis}. Wiley, New York.







\item Chaudhuri, P. (1996) On a geometric notion of quantiles for multivariate data.
{\em J. Amer. Statist. Assoc.}, {\bf 91}, 862-872.

\item Chen, J. and Li, P. (2009) Hypothesis test for normal mixture models: the EM approach. {\em Ann. Statist.}, {\bf 37}, 2523-2542.

\item Chen, H., Chen, J. and Kalbfleisch, J. D. (2004) Testing for a finite mixture model with two components. {\em J. Royal Statist. Soc. Ser.
B}, {\bf 66}, 99-115.

\item Choi, M. and Marden, J. (1997) An approach to multivariate rank tests in multivariate analysis of variance.
{\em J. Amer. Statist. Assoc.}, {\bf 88}, 1363-1370.

\item Claeskens, G. and Hjort, N. L. (2008) {\em Model Selection and Model Averaging}, Cambridge University Press, Cambridge.


\item Dempster, A. P., Laird, N. M. and Rubin, D. B. (1977) Maximum likelihood from incomplete data via the EM algorithm, {\em J. Royal Statist. Soc. Ser. B}, {\bf 39}, 1-38.

\item Duda, R. O., Hart, P.E. and D.G. Stork (2001) {\em Pattern Classification}. Wiley, New York.

\item Dunn, J.C. (1974) Well separated clusters and fuzzy partitions. {\em J. Cybern.}, {\bf 4}, 95-104.

\item Escobar, M., and West, M. (1995). Bayesian density estimation and inference using mixtures.
{\em J. Amer. Statist. Assoc.}, {\bf 90}, 577-588.






\item Gupta, A. K. and Kabe, D. G. (1999) Distribution of Hotelling's $T^2$ and multiple and partial correlation coefficients
for the mixture of two multivariate Gaussian populations. {\em Statistics}, {\bf 32}, 331-339.


\item Hastie, T., Tibshirani, R. and Friedman, J. H. (2009) {\em The Elements of
Statistical Learning: Data Mining, Inference and Prediction}. Springer, New York.


\item Hettmansperger, T. P., Mottonen, J and Oja, H. (1998) On affine invariant multivariate rank tests
for several samples. {\em Statist. Sinica}, {\bf 8}, 785-800.

\item Hettmansperger, T. P. and Oja, H. (1994) Affine invariant multivariate multi-sample sign tests. {\em J. Royal
Statist. Soc., Ser. B}, {\bf 56}, 235-249.

\item Hoeting, J., Raftery, A. and Madigan, D. (1996). A method for simultaneous variable selection and outlier identification in linear regression. {\em Comput. Statist. Data Anal.} {\bf 22}, 251- 271.

\item  Hollander, M. and Wolfe, D. A. (1999) {\em Nonparametric Statistical Methods}. Wiley, New York.

\item Holmes, C. C. and Adams, N. M. (2002) A probabilistic nearest neighbor method for statistical pattern recognition. {\em J. Royal Statist. Soc. Ser. B}, {\bf 64}, 295-306.

\item Jornsten, R. (2004) DDclust clustering and classification based on the L1 data depth.
{\em J. Multivariate Anal.}, {\bf 90}, 67-89.

\item Koltchinskii, V. (1997) M-estimation, convexity, and quantiles, {\em Ann. Statist.}, {\bf 25}, 435-477.

\item Liu, R. and Singh, K. (1993) A quality index based on data depth and multivariate rank tests. {\em J. Amer. Statist. Asssoc.}, {\bf 88}, 252-260.


\item Mottonen, J. and Oja, H. (1995) Multivariate spatial sign and rank methods.
{\em J. Nonparemetr. Statist.}, {\bf 5}, 201-213.

\item Mukhopadhyay, S. and Ghosh, A. K. (2011) Bayesian multiscale smoothing in supervised and semi-supervised kernel discriminant analysis. {\em
Comput. Statist. Data Anal.}, {\bf 55}, 2344-2353.

\item Oja, H. and Randles, R. (2004) Multivariate nonparametric tests. {\em Stat. Sci.}, {\bf 19}, 598-605.

\item Oja, H. (2010) {\em Multivariate Nonparametric Methods with R}. Springer, New York.


\item Puri, M. L. and Sen, P. K. (1971) {\em Nonparametric Methods in Multivariate Analysis}, Wiley, New York.


\item Randles, R. and Peters, D. (1990) Multivariate rank tests for the two-sample location problem. {\em Comm. Statist. Theory Methods}, {\bf 19}, 4225-4238.

\item Rao, C. R. (2001) {\em Linear Statistical Inference and Its Applications}. Wiley, New York.

\item Ripley, B. D. (1996) {\em Pattern Recognition and Neural Networks}. Cambridge University Press, Cambridge.






\item Tibshirani, R., Walther G. and Hastie, T. (2001) Estimating the number of data clusters via the gap statistic.  {\em J. Royal Statist. Soc., Ser. B}, {\bf 63}, 411-423.

\item Titterington, D. M. (1990) Some recent research in the analysis of mixture distributions. {\em Statistics}, {\bf 21}, 619-641.

\end{document}